\documentclass[sigconf, screen]{acmart}
\setcopyright{none}
\renewcommand\footnotetextcopyrightpermission[1]{} 

\newcommand{\licensenotice}{%
  \textcopyright\ 2026 Gene S-H Kim, Nicholas Oskiper, Sean Follmer, and Megan Hofmann.\\This work is licensed under a \href{https://creativecommons.org/licenses/by/4.0/}{Creative Commons Attribution 4.0 International License (CC BY 4.0)}.}

\AtBeginDocument{%
  }
\usepackage{multirow}
\usepackage{subcaption}
\usepackage{cuted}
\usepackage{capt-of}

\begin{document}
\title[Beyond Being Accurate]{Beyond Being Accurate: Implications for Using Natural~Language~Descriptions for Accessible 3D Modeling with~Blind and Low Vision Makers}

\author{Gene S-H Kim}
\orcid{0000-0001-9514-4610}
\affiliation{%
  \institution{MIT CSAIL}
  \city{Cambridge}
  \state{Massachusetts}
  \country{USA}
}

\author{Nicholas Oskiper}
\orcid{0009-0005-3629-7744}
\affiliation{%
  \institution{Khoury College of Computer Sciences,\\Northeastern University}
  \city{Boston}
  \state{Massachusetts}
  \country{USA}
}

\author{Sean Follmer}
\orcid{0000-0001-5592-5949}
\affiliation{%
  \institution{Stanford University}
  \city{Stanford}
  \state{California}
  \country{USA}
}

\author{Megan Hofmann}
\orcid{0000-0003-2283-8587}
\affiliation{%
  \institution{Khoury College of Computer Sciences,\\Northeastern University}
  \city{Boston}
  \state{Massachusetts}
  \country{USA}
}

\renewcommand{\shortauthors}{Kim et al.}
\makeatletter
\AtBeginDocument{%
  \fancyhead[LE]{\@headfootfont\footnotesize\shorttitle}%
  \fancyhead[RO]{\@headfootfont\@shortauthors}%
}
\makeatother

\begin{abstract}
People who are blind or have low vision (BLV) are largely excluded from 3D modeling tools that rely on visual representations of geometry. Natural language descriptions offer a potential path to accessibility, but it is unclear whether language alone can support the deep spatial understanding that editing—not just browsing—requires. Advances in generative AI offer the promise of automatic model description, yet descriptions suited to BLV modelers are structurally distinct from alt-text and from existing 3D model captioning datasets, and the domain-specific requirements for generating them are not well understood. We address this through two coordinated studies: an autoethnographic pilot in which the blind first author worked with 12 sighted participants to develop a taxonomy of descriptive strategies, and a Wizard-of-Oz main study that applied those strategies as idealized scripts with 10 BLV participants. We identify a systematic \textit{use-to-editing gap}: descriptions that support accurate but incomplete mental models are insufficient for specifying modifications, because editing requires knowing the state of features the user did not know to ask about. We show that this gap produces four failure types—accuracy, completeness, expression, and navigation—that accuracy-focused evaluation frameworks cannot detect. We propose a two-phase evaluation framework that diagnoses the failure type a description system exhibits and guides developers toward targeted design interventions, thereby providing a methodological standard for assessing whether description-based systems genuinely support BLV modelers as creators.
\end{abstract}

\keywords{Accessibility, Blind, Low-vision, Visual Impairment, Visual Access Technologies, Generative AI, LLM, 3D Modeling,  3D printing, Digital Fabrication, Making}

\maketitle
{\footnotesize\renewcommand\thefootnote{}\footnotetext{\licensenotice}}

\newpage
\section{Introduction}

Design and engineering have been transformed by computer-aided design (CAD) tools for 3D modeling. Unfortunately, people who are blind or have low vision (BLV) are excluded from these tools, because 3D modeling relies on visual representations of geometry that are difficult to translate into accessible modalities (e.g., text, audio, tactile). Sometimes a model can be 3D-printed or displayed on a haptic display (e.g.,~\cite{Siu_shapeCAD}) for tactile exploration. However, this severely limits BLV users' ability to lead their own modeling tasks. BLV users may instead rely on text descriptions, but this raises the question:
\textit{are visual descriptions sufficient for comprehending 3D models for design tasks?}

If visual descriptions are sufficient, then emerging generative AI and large language models (LLMs) offer the promise of automatic model description. These systems can generalize across domains and translate images into natural language, and their capabilities are advancing rapidly. They are already being applied to support 3D model exploration~\cite{zhang_a11yshape_2025} and other forms of visual content (e.g.,~\cite{editscribe, altcanvas}). However, generative AI exhibits unique behavior when applied to novel domains with little or no corpus of well-suited training data (e.g., 3D modeling~\cite{3D_Generative_AI}). Visual descriptions of 3D models that support BLV modelers will be distinct from alt-text descriptions, the documentation supplied by makers (e.g., on Thingiverse~\cite{thingiverse}), or labels and captions used to create datasets for virtual applications such as animation (e.g.,~\cite{Xiang_objectnet3D}). Tailoring an LLM to generate these descriptions will require a deep understanding of what information is needed and how best to present it to BLV modelers.

Determining the value of visual descriptions as a method for accessible 3D modeling and developing implementation guidelines for AI-based description systems poses unique methodological challenges. Foremost, we cannot rely on existing information on accessible 3D modeling by BLV people, as these practices are extremely rare and often rely on individualized, interdependent design practices co-created by BLV modelers and sighted assistants. Further, we cannot rely on prototype AI systems at these early stages of work because these systems often exhibit novel and unrepeatable behaviors (e.g., hallucinations~\cite{hullucination_mitigation}). Additionally, it is unclear if the 3D model representations suited to building these AI systems~\cite{3D_Generative_AI} can retain the features BLV modelers require.

Considering these challenges, we conducted two coordinated studies that draw on Kim's lived experience as a Blind engineer who uses visual descriptions to create and edit 3D models. In an \textit{autoethnographic pilot study}, we recreated this workflow by having 12 sighted participants describe 3D models to Kim as he formed clay replicas to express his understanding of each model's geometry. From these sessions, we surfaced a taxonomy of descriptive strategies that sighted assistants use when describing 3D models to BLV modelers. For our main study, we operationalized these strategies as scripts and conducted a \textit{Wizard-of-Oz study}~\cite{Dahlback_WizardofOz_why_and_how} in which sighted researchers used the scripts to act as AI agents describing models to 10 BLV participants. We analyze these conversations and the clay replicas to understand the potential benefits and limitations of visual-description-based 3D modeling tools.

From these studies, we identify a systematic gap between the level of model understanding currently supported by visual descriptions and the level required for editing. Sighted assistants in the pilot encountered significant challenges generating sufficient geometric detail and resolved them by establishing a shared frame of reference and common language~\cite{clark1991grounding, resnick1991perspectives} with the BLV modeler. BLV participants in the main study found descriptions adequate for forming a general impression of models but consistently inadequate for specifying modifications---revealing what we term the \textit{use-to-editing gap}. 
We discuss these findings in terms of three criteria that descriptions must satisfy to support model editing: they must be \textit{complete}, \textit{expressive}, and \textit{navigable}. Failures in each dimension produce characteristic error patterns that accuracy-focused evaluation frameworks cannot detect.

Based on these findings, we propose a two-phase framework for evaluating description systems for 3D modeling tasks. The framework diagnoses which type of failure a system produces and directs developers toward targeted design interventions. We conclude that visual descriptions alone are unlikely to be a sufficient medium for accessible 3D modeling, but that this does not preclude a meaningful role for generative AI in this domain when systems are designed with the use-to-editing gap explicitly in view.

\section{Related Work}

Many researchers have demonstrated the value of 3D printing for spatial design tasks~\cite{3D_print_for_editing_web_spatial_layout}, accessible education~\cite{investigating_imp_3d_printing_sped, digital_fabrication_for_accessible_education_content_producers, 3d_print_for_stem_education_acess}, creating tactile maps~\cite{maptimizer, 3d_print_v_tactile_maps}, and deeper understanding of real-world objects through interactive 3D models~\cite{Designing_Interactions_for_3D_Printed_Models_with_Blind_People}. Research on how BLV people design and modify their own assistive technology reveals that blind people already engage in sophisticated customization practices, combining, hacking, and switching between tools and artifacts to meet needs that off-the-shelf solutions do not address~\cite{herskovitz_hacking_2023}. Despite the clear advantages of 3D printing~\cite{accessible_maker_space_blaser, SCAFFOLDing_Young, making_making, accessible_fab_weaving, accessible_EE_education_blind}, 3D modeling remains an open accessibility challenge.

\subsection{Accessible CAD}

3D modeling tools require visual interaction and are not screen reader accessible. Instead, BLV modelers use programmatic CAD systems to create 3D models~\cite{jscad, SolidPython, openSCAD, jscad-guy}. Additionally, there are text-to-3D-model generative AI systems~\cite{more_text_to_3D, Sella_Vox-E_Text-Guided_Voxel_Editing_of_3D_Objects_ICCV_2023_pape} that have accessible inputs but still leave the BLV user with an inaccessible output.

Prior work has explored accessible representations of 3D models using haptics (e.g.,~\cite{Archi-TangiBlock, Siu_shapeCAD, haptic_mesh_insepection_for_blind_modelers}) and audio (e.g.,~\cite{shapesonic, haptic_mesh_insepection_for_blind_modelers, zhang_a11yshape_2025}). These include ShapeCAD~\cite{Siu_shapeCAD}, a 2.5D shape-changing display that interfaces directly with OpenSCAD. While effective, it is limited by its comparatively low resolution. Other low-cost approaches have been explored: Thormahlen et al. developed a one-end-effector device that enabled BLV people to feel and hear sonified representations of 3D objects~\cite{haptic_mesh_insepection_for_blind_modelers}, and audio alone has been explored as a medium~\cite{shapesonic}. However, these approaches cannot convey many complex features of 3D models (e.g., concavities, smaller segments, precise contours). A widely available, domain-agnostic solution for non-visual 3D modeling has yet to be fully realized.

Beyond 3D models, editing and creating visual content are important focuses of accessibility research, and emerging generative AI technologies may enable novel solutions~\cite{digital_content_creation_needs}. GenAssist~\cite{genassist} makes text-to-image creation more accessible by empowering users to verify generated details, obtain information not included in the prompt, and compare image candidates. To support the iterative nature of image creation, EditScribe~\cite{editscribe} provides multiple types of natural-language verification, enabling independent image editing. AltCanvas~\cite{altcanvas} enables blind people to construct visual scenes incrementally using a tile-based interface with speech and audio feedback. Applying similar approaches to 3D modeling, A11yShape~\cite{zhang_a11yshape_2025} integrated LLM-generated descriptions with an OpenSCAD programming interface, enabling BLV programmers to create and verify 3D models through synchronized representations of code, structure, and AI-generated textual feedback. Our work more deeply explores how BLV users form mental models of complex 3D geometries via natural language, especially in contexts where users may not have access to well-decomposed, hierarchical openSCAD code.

\subsection{Natural Language for Accessibility}

Natural language has successfully been used to represent visual information in other contexts. The Web Content Accessibility Guidelines~\cite{wcag} require text alternatives for visual media, and researchers have conducted extensive work to improve the quality and availability of alt text for images. These include datasets with long-form answers to BLV users' visual questions~\cite{long_form_answers_viswiz}, tools supporting humans who write alt-text~\cite{supporting_quality_alt_text_writers}, and improving AI-generated alt-text~\cite{diagram-input-LLM-for-STEM, be_my_ai}. To the best of our knowledge, no prior work has articulated design standards for writing alt-text specific to 3D models. While not explicitly motivated by accessibility, methods to automatically caption 3D models exist~\cite{Chen_2023_ICCV, cap3D}; however, upon examining these datasets and piloting them, Kim found these captions inadequate for providing the geometric information required for 3D modeling.

A growing body of work demonstrates that the context in which an image description is used substantially shapes what information it must contain~\cite{stangle_whatVIwant, Stangl_Beyond_One_Size_Fits_All}. BLV users' information needs vary across sources, tasks, and prior knowledge, and context-aware descriptions that account for surrounding content better satisfy those needs~\cite{context_aware_image_descriptions}. These findings motivate a distinction this paper formalizes: the requirements of a description intended to support model selection differ structurally from those of a description intended to support model editing.

Effective descriptions of 3D objects must also navigate the cognitive demands of spatial comprehension from language alone. Research in spatial cognition establishes that people construct mental models of described environments by organizing spatial information around extensions of body axes~\cite{franklin_tversky_searching_1990}, and that speakers systematically adopt different reference frames---intrinsic, relative, and absolute~\cite{levinson_space_language_2003}---when describing spatial relationships, routinely alternating between frames rather than committing to one throughout~\cite{taylor_tversky_perspective_1996, schober_perspective_taking_1993}. For BLV users, who must construct spatial mental models of novel objects without visual reference, this variability has particular consequences: Noordzij et al.~\cite{noordzij_spatial_mental_models_2006} found that route-based (egocentric) framings helped BLV participants form spatial mental models from verbal descriptions. 

\subsection{Generative AI for Accessible Image Modification and Verification}
Generating effective visual descriptions of complex 3D content is a nuanced task. Generative AI may enable on-demand access to visual descriptions at scales far beyond the limitations of sighted assistants; however, scale will not reduce BLV users' need for high-quality descriptions.

Alongside these promising capabilities, an emerging body of work documents that BLV users face a fundamental verification challenge with AI-generated visual content: they cannot independently confirm whether AI outputs are accurate. Alharbi et al.~\cite{missfitting_ai} found that blind users develop a range of compensatory verification strategies, but that errors remain difficult to detect without visual access. This \textit{verification asymmetry} is compounded when AI systems rely on inaccessible visual explanations (e.g., heatmaps)~\cite{object_recognition_errors}. Research on BLV users' mental models of AI systems reveals that they often develop flawed understandings of how these tools work, sometimes treating AI as more authoritative and reliable than it is, which increases the risk of undetected errors in AI outputs~\cite{BLV_GenAI_ASSETS24}. These tendencies are not failures of user judgment but consequences of systems that do not support accessible error detection. Autoethnographic research by researchers with disabilities confirms that when AI is used for accessibility tasks, the source material, the AI output, or the verification mechanism is often itself inaccessible~\cite{Glazko_Autoethnographic_Ai}.

These verification challenges have motivated work to make AI uncertainty legible and help BLV users develop appropriate skepticism. ImageExplorer~\cite{image_explorer} demonstrated that multi-layered tactile exploration of images, rather than reading AI-generated captions, led blind participants to develop greater skepticism toward AI errors. Ehsan et al.'s~\cite{Ehsan_Seamful_XAI} framework of Seamful XAI argues that systems should strategically reveal their limitations rather than concealing them, enabling users to adapt their reliance on AI accordingly. Together, these findings establish that description systems for BLV users must not only generate accurate content but must also communicate their limitations to BLV users.

\section{Methods}
\begin{table*}[t]
\centering
\caption{Summary demographics for BLV-modeler participants.}
\label{tab:modelers}
\vspace{-10pt}
\begin{tabular}{@{}llllrl@{}}
\toprule
\textbf{ID} & \textbf{Gender} & \textbf{3D Modeling Experience} & \textbf{Relevant Experience} & \multicolumn{1}{l}{\textbf{Models}} & \textbf{Study Location} \\ \midrule
M1  & Male   & None                    & Technical Experience & 1,2   & Remote                        \\
M2  & Female & None                    & No Experience        & 3,4   & Northeastern University     \\
M3  & Female & None                    & Technical Experience & 5,6   & Remote                        \\
M4  & Male   & When Sighted            & Technical Experience & 7,8   & Remote                        \\
M5  & Male   & While having Low-Vision & Technical Experience & 9,10  & Northeastern University      \\
M6  & Female & None                    & No Experience        & 13,14 & [Anonymized Blindness Center] \\
M7  & Female & None                    & No Experience        & 15,16 & [Anonymized Blindness Center] \\
M8  & Male   & None                    & No Experience        & 17,18 & [Anonymized Blindness Center] \\
M9  & Female & None                    & No Experience        & 19,20 & [Anonymized Blindness Center] \\
M10 & Male   & None                    & No Experience        & 11    & [Anonymized Blindness Center] \\
\bottomrule
\end{tabular}
\vspace{-14pt}
\end{table*}

We conducted two coordinated studies following a research through design approach~\cite{Zimmerman_Research_through_Design} grounded in autoethnographic methods~\cite{Glazko_Autoethnographic_Ai, hofmann_living, accessible_EE_education_blind}. First, we conducted an \textit{autoethnographic pilot study} in which Kim collaborated with sighted participants to surface the descriptive strategies that sighted assistants use when describing 3D models. 

Based on the results of this pilot, we then conducted a main \textit{Wizard-of-Oz study} in which 10 BLV participants explored 3D models using descriptive scripts read by a sighted assistant. The model exploration procedure was identical across both phases, with participant roles reversed. In the pilot, Kim served as the BLV modeler, and sighted participants served as assistants. In the main study, BLV participants served as modelers, and sighted researchers used scripts to simulate an AI assistant's role.

In both the pilot and the main studies, the sighted assistant was given a brief standardized description of the model (\autoref{app:3D_Model_Descriptions}) and a link to an online 3D model viewer~\cite{3Dviewer}. After hearing the model description, the BLV modeler asked the assistant open-ended questions to elicit further details about the model. Participants in the pilot study provided free-form responses, while researchers in our main study were restricted to responses included in the scripts. Based on the sighted assistant's answers, the modeler recreated the 3D model in modeling clay, which we use to assess comprehension. We shipped modeling clay to remote participants in advance. During construction, the clay was hidden from assistants to prevent them from influencing descriptions. This restriction serves two purposes: we did not want assistants to intervene or correct the modeler, as this would strongly influence comprehension; and, to evaluate the potential utility of AI assistants, we presume that AI systems would not have access to information about the BLV modeler's current understanding beyond what is expressed in language. All participants explored both models except M10, who reached saturation on their first model in 50 minutes and did not have time to proceed to the second.

After modelers were satisfied with their clay replica or determined they could not gain more information, they reported their confidence in their understanding on a 7-point Likert scale. These ratings are not used as measures from which we derive empirical claims; rather, they prompted modelers to articulate their sense of understanding before transitioning to the post-session interview. Next, we asked BLV modelers to make a single design modification to the model as if they were prompting an LLM to do so. We use these prompts to further contextualize the modeler's comprehension following the visual description process (e.g., how specifically can they describe the requested changes?). After each study session, we concluded with a semi-structured interview asking participants about challenges, strategies, and overall reflections. Sessions lasted up to one hour.

Because no standardized method for describing 3D models exists, and no existing datasets of descriptions suited to BLV modelers are available (e.g., existing 3D model captions~\cite{luo2023scalable, luo2024view} are not designed for visual description), our pilot could not rely on prior resources. Instead, it required a BLV modeler who could both anchor the interaction and contribute expert analytical judgment. Kim was uniquely suited to fill this role. He was born blind, maintained roughly 20/400 vision until age 15, and has since had only light perception and color contrast detection. His prior experience with 3D modeling using conversational visual description informed our goal of exploring AI-generated visual descriptions as a medium for accessible 3D modeling and motivated the format of both studies. Unlike prior work that simulates conversational assistant workflows~\cite{Salisbury_Conversational_Crowdsourcing}, this pilot does not require simulation of disability. 

The autoethnographic dimension of the pilot lies in this dual role: Kim participated as the BLV modeler whose comprehension shaped each interaction, and simultaneously served as an analyst whose lived experience with non-visual 3D modeling informed the identification and interpretation of sighted assistants' descriptive strategies. The pilot findings accordingly report on those strategies rather than on Kim's experience in isolation, because the strategies are what we needed to operationalize to conduct our main study with BLV participants.

\subsection{Data Collection and Analysis} \label{sec:analysis}

Across both phases, we collected audio transcripts, screen recordings, images of the clay replicas, and self-reported confidence ratings. We analyzed this data using a reflexive thematic analysis. Kim and Hofmann independently coded half of the sessions inductively, then discussed their observations and developed themes. Additional data was referenced to contextualize codes or clarify transcripts. For example, Hofmann, a sighted author, referenced screen recordings of the pilot sessions to ensure they understood which parts of a model were being described at each moment. The pilot study data served a dual purpose: contributing findings about sighted assistants' descriptive strategies and directly informing script development for the main study.

\subsection{Participants} \label{sec:participants}

We recruited ten BLV participants for our main study (\autoref{tab:modelers}), and twelve sighted participants for our pilot study (see~\autoref{app:sighted_participants}). All participants were recruited from our personal networks and through snowball sampling, provided unsigned verbal consent, and were compensated with their choice of a \$50 digital Visa or Amazon gift card. Northeastern University's IRB approved all procedures.

We required BLV participants to identify as having a visual impairment that made 3D modeling tools inaccessible, and we prioritized recruiting those with 3D modeling or engineering experience. Despite these efforts, only two BLV participants had prior experience with 3D modeling, both of whom also had relevant domain expertise; two additional participants had programming experience. 

We classified sighted participants as having sighted-assistant experience if they had prior experience creating alt-text or providing visual descriptions for BLV people, and as having domain expertise if they had a bachelor's degree or were pursuing one in an engineering-related field.

We argue that it is valuable to study both experienced and novice modelers across both participant populations. Visual 3D modeling tools are critical for engaging in expert-amateur-making activities~\cite{Kuznetsov_DIY} and have been designed to serve both novice and expert users. 
Sampling participants across the spectrum of experience also reflects the realistic conditions under which AI description systems would be deployed: both the sighted assistants who inform such systems and the BLV modelers who use them will span a range of prior knowledge. If the descriptive strategies surfaced in the pilot do not serve either experienced or novice BLV modelers, they are unlikely to form a sufficient basis for accessible 3D modeling tools.

\subsection{3D Model Sampling}

To capture real-world variation in visual descriptions across model types and contexts, we systematically selected a diverse set of models from the popular 3D printing repository Thingiverse~\cite{thingiverse} (see \autoref{app:models}).  Since there are no existing non-visual 3D modeling tools, we chose to use pre-existing models, presumably created by sighted modelers, to ensure sufficient complexity for comparison with real-world modeling tasks. Since model reuse is critical for teaching novice modelers~\cite{Kuznetsov_DIY}, any non-visual 3D modeling tool must support models of this complexity rather than expecting blind modelers to always create models from scratch.

We used stratified random sampling to select models from the 1,000 most popular models in June 2024. To ensure variation would prompt unique descriptions, we stratified by four model categories: figurines, anatomical/scientific, models of man-made objects, and functional components. These categories were derived from Kim's reflexive thematic analysis of Thingiverse models. They are not comprehensive nor independent; however, they diversified the dataset while systematically capturing models relevant to BLV modelers. In each session, participants were given two unique models, one from each category, with equal numbers of participants assigned to each category. The research team generated brief standardized descriptions used to introduce each model.

\subsection{Autoethnographic Pilot Study} \label{sec:pilot}

The autoethnographic pilot surfaced a consistent structural approach across all sighted participants and a taxonomy of expressive strategies that varied across participants and model types. Together, these findings directly informed the content and structure of the main study scripts. 

From the assistant's perspective, describing 3D models goes beyond creating alt-text. Five of the six participants with sighted-assistant experience noted, without prompting, that describing 3D models was inherently different and far more challenging. A6
said: ``\textit{This was a lot more challenging than writing alt-text\ldots I had to get things relationally as accurate as I could\ldots Alt-text for an image, it's more of a 2D environment. You can just look at the context.}'' Context was the critical difference; 3D models have far more complexity with far less context, which introduced novel challenges that do not arise in standard alt-text workflows.

\subsubsection{Segmentation and Shared Language} \label{sec:shared_language}

We observed a common structural approach across all 12 sighted participants: they began by providing an overview of the model, then segmented it into labeled sub-components that served as reference anchors throughout the conversation. Once shared vocabulary was established, assistants consistently returned to these labels rather than redescribing geometry from scratch. Examples include ``\textit{stairs}'', ``\textit{door}'', and ``\textit{wall}'' for buildings (A5, A7, A9, A11); ``\textit{arms}'', ``\textit{legs}'', and ``\textit{nose}'' for figurines (A1, A2, A3); and ``\textit{thread}'', ``\textit{joint}'', and ``\textit{gear}'' for engineering components (A2, A4, A8, A10, A11).
When conventional labels were unavailable (e.g., due to a lack of shared terminology or background expertise), participants adopted one of two strategies: pure geometric descriptions (e.g., ``\textit{the upside down pyramid}'' (A12), ``\textit{the semisphere}'' (A9)) or simile approximations that drew on everyday objects (e.g., ``\textit{shape of an iPhone, but thinner}'' (A1), ``\textit{looks like an ice cream cone}'' (A2), ``\textit{a belt of windows around the tower}'' (A11)). 

After establishing individual segments, assistants related them to one another spatially, using relative comparisons of shape, orientation, and position to build up a picture of the whole model (A1, A2, A3, A5, A6, A7, A8, A10). As A2 reflected: ``\textit{I think what worked well is this high-level to detail approach.}''
This segmentation-first pattern directly shaped the structure of the main study scripts: each script opens with a model overview and then organizes the description by component, mirroring the approach that all pilot participants independently converged on.

\subsubsection{A Taxonomy of Expressive Strategies} \label{sec:descriptive_stategies}

Beyond the shared structural approach, sighted assistants deployed a variety of expressive strategies to convey geometric detail. 

\paragraph{Geometric Descriptions.}
All participants relied on geometric terms (e.g., ``\textit{rectangular prism}'') to describe at least one model feature. Many supplemented these with mathematical language (e.g., ``\textit{symmetrical}'', ``\textit{parallel}'', ``\textit{bisect}'') and relative dimensions (e.g., ``\textit{twice as long as the base}'' (A7)). For complex shapes, participants described geometric transformations: ``\textit{It's a circle except you pull the two halves apart and extend it}'' (A2); ``\textit{It's exactly the same as the first U, but instead, it's flipped vertically 180 degrees}'' (A10).

\paragraph{Comparisons.}
All participants relied on comparisons to common objects or symbols: ``\textit{A little bit flatter than a carrot typically would be}'' (A7); ``\textit{It's the capital letter E, and then directly next to it is the capital letter L}'' (A4). When comparisons did not resonate, assistants fell back on geometric descriptions: ``\textit{like a triangular prism, but curved}'' (A12).

\paragraph{Embodied Metaphors.}
All participants used embodied metaphors and personification, particularly to describe figurine poses or anthropomorphic models. For example, A3 described the Statue of Liberty: ``\textit{Her left hand is holding a book. Her elbow is down, and her hand is up.}'' Embodied metaphors extended to inanimate objects as well, with both the model and the modeler as the embodied subject: A4 said ``\textit{As if you're holding an umbrella and closing the umbrella.}'' Some hypothetical metaphors framed the model as something to be constructed: A10 said ``\textit{I would start putting the cylinder underneath the large size ball, and then pinch off a little bit of the clay and move it up until it touches the ball.}''

\paragraph{Reference Frame Variability.}
Participants naturally alternated between reference frames without consistency. Some anchored descriptions to the modeler's body and grip (e.g., ``\textit{You're holding it in your hand like you'd hold a cup}'' (A4)), while others used traditional viewing perspectives (e.g., ``\textit{top view}'', ``\textit{profile view}'') or the object's own axes. This variability is consistent with research showing that effective spatial description often requires active frame coordination rather than a fixed perspective~\cite{schober_perspective_taking_1993, taylor_tversky_perspective_1996}. No participant remained consistently committed to a single frame; all shifted among intrinsic, relative, and embodied orientations over the course of a session.

This variability motivated a key script design decision: each script includes multiple framings for each component geometry, so that at least one is likely to align with a given BLV modeler's existing spatial knowledge.

\subsection{Sighted-Assistant Script Development} \label{sec:script_development}

After analyzing descriptive strategies used by sighted assistants in our pilot, we transitioned to studying BLV modelers. To do this, we controlled the descriptive strategies delivered by sighted researchers using prepared scripts. For each model, we began by identifying all components (e.g., body parts, salient features) described by the sighted participant from the pilot who had described that model. We then structured the script with an overview section and a per-component section for each identified part. These sections were populated with direct quotes from the pilot session, each labeled with the expressive strategy it represented. We then identified which strategy types were absent from each section and wrote additional descriptions for those components to fill the gaps. Where new descriptions were needed, a sighted author used ChatGPT to generate candidate phrasings that mimicked the expressive patterns from our pilot study data. The sighted researcher verified all generated descriptions to ensure that they accurately represented the models. This approach produced comprehensive scripts that covered every component identified by pilot participants, with each component represented by every expressive strategy.

These scripts represent an idealized description agent: complete in component coverage, consistent expressive range, and free from hallucination. This is a deliberate methodological choice that isolates the question of whether the description \textit{medium} supports BLV modelers from the separate question of whether any existing AI system is reliable enough to deliver descriptions of sufficient quality. The capabilities and limitations of generative AI technologies are rapidly shifting; before building systems, we must first determine whether the medium of visual description serves BLV modelers' needs at all.

\section{Findings} \label{sec:findings}

We structure our findings around three themes: an emergent tension between providing sufficient detail and overwhelming modelers with excessive description length; modelers' varied approaches to eliciting and navigating descriptions; and the gap between model comprehension and the specification of model edits.

\subsection{Overwhelming Description Depth}

The conversational structure of these interactions introduces complexities not present in systems that rely solely on alt-text. Additional information can be requested, generated, clarified, and revisited. However, we observed a consistent trade-off. Descriptions were often too lengthy and complex to interpret, yet all details were essential---they were simultaneously too long and too short.

\subsubsection{Overwhelming Description Length}

All participants found the overview descriptions overwhelming. Participants either interrupted the assistant while the overview was being read aloud (M2, M4, M5), asked for the overview to be repeated in its entirety (M1, M8), or asked for a repeat description of a specific section of the overview (M3, M7, M10). Some participants articulated the challenge of keeping track of details in lengthy overviews. As M10 put it: ``\textit{I couldn't process everything that was going on. Nobody can listen and create a perfect model unless you're superhuman.}'' Participants also noted how natural language descriptions felt unequal to a sighted modeler's experience due to the sheer quantity and length: ``\textit{With language, it takes a lot of description to include every single detail. It may take me three pages to get something that someone sighted might get in 15 seconds.}'' (M5).

While no participants wanted details removed, they found irrelevant information particularly irritating. M4 stated, ``\textit{The whole detailed description about the eye sockets didn't make any sense...It's a skull! That whole description was completely useless to me.}'' M8 expressed related frustrations: ``\textit{I have a general idea of what it's supposed to be, but the words `intricately detailed' seem vague.}''

\subsubsection{The Value of More Details}

Despite comments and behaviors indicating that the descriptions were lengthy and overwhelming, nearly all participants wished for more details in the script. Many participants asked multiple questions about the same segment and/or requested multiple descriptions of it (M2, M3, M4, M8, M10). Some participants requested details not present in the script (e.g., the contours of claws (M1), the textures of features (M8), and the relative dimensions of specific parts of the model (M3, M4, M10)). When details were missing, participants often expressed their frustration. For example, after the researcher could not provide a specific count of the number of polygons on the cranium of the low-poly skull, M4 responded with frustration: ``\textit{That's surprising. That would be a common question to ask for anyone dealing with a polygonal representation of something.}'' This detail is particularly salient when considering the encoding of a 3D model. Perhaps counting polygons is trivial when encoding a 3D mesh, but this may be impossible when using other representations (e.g., shape-descriptors, volumetric data, or point-clouds)~\cite{survey_3D_Data_Rep_for_DL}.

Participants with prior modeling and/or technical experience tended to be more specific about the type of information they were seeking (e.g., M4 asking for polygon counts; M3 and M4 asking for specific dimensions). However, participants without expertise also asked for this level of detail (e.g., M10, M8). Our participants represent a wide diversity of experiences and backgrounds beyond our classification of 3D modeling and technical expertise. It is likely that each individual's preference for specific details and tolerance for abundance will vary depending on context.

A particularly important form of context is what the person intends to do with the 3D model and what they need to know to enact their design intent. It was common for participants to realize they were missing details in their understanding of the 3D model only after attempting to describe design modifications. The majority of participants (M1, M2, M3, M5, M7, M10) expressed confusion about which modification to make because they realized they did not know whether the model already had features that might clash with their proposed changes. Two examples worth highlighting are limbs of figurines: ``\textit{I don't know if the picture had it. Dragons have sharp claws... I think I'll add that to the model...mine doesn't have [claws].}'' (M1); and non-tangible features on surfaces: ``\textit{I don't think it mentioned any writing on the building. I'm sure it is something like `United States Supreme Court'...I'd inquire about those details.}'' (M5).

\subsection{Approaches to Navigating Descriptions}

Participants used various strategies to manage the tension between overwhelming details and missing details. Access to a variety of descriptive techniques appears to be critical to model comprehension. Generally, participants expressed that having multiple different descriptions would help clarify their understanding: ``\textit{A variety of ways to describe it could have been more helpful. Even those nitty-gritty things.}'' (M2).

Many participants expressed the benefit of descriptions that combine and freely transition between multiple description styles (M2, M3, M4, M8, M10). For example, participants expressed that embodied metaphors complemented geometric descriptions. As M4 put it: ``\textit{if I said, `I don't think I fully understand this model. Can you give me a step-by-step guide to build this with simple shapes?' That would be extraordinarily helpful.}'' Rather than preferring one style of description, participants found that a diversity of description styles increased the likelihood that one would resonate and fit their understanding of geometry.

When descriptions were insufficient, most participants (M1, M2, M3, M5, M7) reported that it was challenging to develop the right questions to elicit sufficient detail about the model. Even individuals like M1, who has a significant technical background, expressed that it was challenging to prompt for details they needed: ``\textit{I was confused on how to actually shape this...But when I was making it, I wasn't sure how to ask those questions.}'' When participants struggled to elicit a new way of describing details, a common strategy was to present a hypothetical description of the feature to test whether they could accurately describe the model (M1, M2, M3, M5, M10). 

For example, ``\textit{Does it mean that they are attached to each other, or is the tail really like tails attached to the back of the animal?}'' (M5). In these cases, the assistant would not deviate from their script and did not answer directly. Instead, they repeated relevant phrases from the script or, if there were no similar phrases, stated that they could not provide an answer.

While variety facilitated understanding specific details, most participants benefited from a consistent model overview. For instance, M3, M5, and M10 appreciated the detailed overview. As M5 put it, ``\textit{I like getting a bigger picture of the overall style and then getting more into shape and specific features. I thought that was genuinely effective.}'' For some participants (M4, M10), using the overview to navigate the model made it easier to describe the desired modification. M10 explained, ``\textit{I was using multiple frames of reference...The point is that if it went where it was supposed to go, then you'd have three to four different reference points.}''

\subsection{Challenges Specific to Model Editing}

While many participants demonstrated an advanced understanding of the 3D models, many still struggled to translate that understanding into specific modifications to the models. This often resulted in vague prompts for substantial changes. For example, participants prompted modifications to move an existing feature (M10), transform a feature (M3, M5, M7), change the surface (e.g., texture/color) (M1, M3, M5, M8), and add a new feature to the model (M1, M2, M4). In some cases, the modification would fundamentally change the model's geometry and functionality or require generating an entirely new design. For example, M2 said, ``\textit{Could you modify this model [to] attach to the knee joint instead of to the hip?}'' As an example of functional change, M7 asked, ``\textit{Make it taller underneath, so the cover will be higher, but it'll still protect the wires.}'' Similarly, M4 stated, ``\textit{Provide me with the same model with an inner ring that connects all four outer gears.}'' Even seemingly simple changes required multiple smaller changes across the model that participants could anticipate but not fully articulate (M1, M5, M7). For example, M7 asked about a prosthetic hand, ``\textit{Make a bigger hand. Make [the] space bigger between each finger.}'' They recognized size's effect on spacing but could not describe the relationship in detail.

Some participants reported that, due to the sheer amount of detail required to understand the model's features, they made many implicit assumptions about it and realized they needed more information only when attempting to modify its design. That is, a different level of detail was desired when editing the model compared to simply understanding it. As M3 put it, ``\textit{So in terms of image description, it was great. But in terms of me actually needing to create a model, I felt like there was some information...that I really need.}''

\section{Discussion}\label{sec:discussion}

Our findings revealed that using natural language to describe 3D models for BLV modelers presents a categorically different challenge from creating alt-text for images, one that cannot be adequately assessed by accuracy metrics alone. Sighted assistants consistently noted the difficulty of generating sufficient detail, while BLV participants found descriptions inadequate for comprehending models sufficiently to edit them. We found a systematic gap between the level of understanding descriptions supported and the level that editing requires. This finding motivates a reexamination of how description systems for BLV users are evaluated. Prior frameworks have largely treated description quality as a single-dimensional property, assessed by accuracy as judged by sighted evaluators or by how well a description serves BLV users as consumers of visual content, not as creators of it~\cite{supporting_quality_alt_text_writers, stangle_whatVIwant, Stangl_Beyond_One_Size_Fits_All}.

Beyond accuracy, we present additional criteria that descriptions must satisfy to enable model editing. First, descriptions must be \textit{complete} and provide all details needed for the BLV user's task. Second, descriptions must be \textit{expressive} and present information in multiple consistent ways. Third, descriptions must be \textit{navigable} and scaffold the user's interaction, enabling them to access all the details relevant to their task. The first criterion encompasses two structurally distinct failure types: an \textit{accuracy failure}, in which a description contains factually wrong content, and a \textit{completeness failure}, in which it omits relevant details. \textit{Expression failures} and \textit{navigation failures} correspond to the second and third criteria, respectively. Each failure type both limits model use and prevents crossing the gap to editing.

\subsection{Complete Descriptions for Use and Editing}

A standard approach to evaluating AI captioning systems is to measure accuracy and error rates against ground truth established by sighted evaluators~\cite{supporting_quality_alt_text_writers}. However, high accuracy may mask omissions of critical details needed to verify a BLV modeler's understanding of the model or to edit it effectively. Prior work has explored the \textit{verification asymmetry}~\cite{missfitting_ai} between BLV users and sighted evaluators. While a sighted evaluator assesses description accuracy by cross-referencing it with visual perception, a BLV user has no independent verification. As a result, even rare or small errors can create misconceptions that human assistants struggle to correct~\cite{Salisbury_Conversational_Crowdsourcing} and these misconceptions may propagate silently into the BLV user's downstream reasoning~\cite{stangle_whatVIwant, context_aware_image_descriptions}.
These consequences emerged at the transition from comprehension to editing. Participants frequently formed confident spatial mental models during exploration. Yet a consistent pattern emerged: gaps in participants' understanding that were invisible during exploration became consequential at the moment of editing (e.g., M3's sense that information was missing, M1's assumption that the dragon figurine did not have claws).

We call this the \textit{use-to-editing gap}: descriptions that support accurate but incomplete mental models of a 3D object do not necessarily support specifying modifications to that object. Use requires enough information to form a correct global impression; editing requires knowing the precise state of any feature that a proposed modification might interact with, including features the user did not know to ask about.

Two distinct failure types arise from inadequate description content. An \textit{accuracy failure} occurs when a description contains factual errors that generate misconceptions about what a segment contains or how it is shaped. These errors are particularly harmful because the verification asymmetry means BLV users cannot detect them independently; an inaccurate description and a correct one are indistinguishable without visual access to the model. 
A \textit{completeness failure} occurs when a description omits details that, while not incorrect in what they say, leave the BLV user unaware of features that are present in the model and relevant to their task. Completeness failures are structurally distinct from accuracy failures: a description can be entirely accurate in what it says, yet still too incomplete to support editing.

\subsection{Expression: Aligning Frames of Reference}

Beyond assessing the content of descriptions, an effective description for 3D modeling requires attention to how that content is expressed. No single expressive strategy reliably supported spatial understanding across BLV participants or model types---this was not a matter of individual preference alone.

One explanation is that spatial descriptions necessarily adopt reference frames that make some properties salient and others invisible. Levinson's~\cite{levinson_space_language_2003} intrinsic, relative, and absolute reference frames each highlight different spatial relationships; no single frame conveys what the others do. Sighted assistants in our pilot study naturally alternated between these frames, some orienting the model relative to the modeler's body and grip, others describing a figure's features using its own limb axes and pose. Because mental models of described environments are organized around body axes~\cite{franklin_tversky_searching_1990}, descriptions in different frames map differently onto the user's spatial representation, and effective spatial description often requires active frame coordination between interlocutors~\cite{schober_perspective_taking_1993}. Access to multiple framing strategies increases the likelihood that at least one will align with the BLV user's existing spatial knowledge.

Multiple expressive strategies also function as a verification mechanism. When two strategies describe the same feature and produce the same spatial inference, their consistency provides evidence that the description is correct, even without direct visual verification. When they produce different inferences, the inconsistency is a detectable signal of ambiguity or error. BLV participants spontaneously exploited this, presenting hypothetical descriptions of a feature to test against the researcher's response, using it as a consistency-testing strategy that enabled verification without visual access. Description systems that structurally encourage this behavior by proactively offering multiple framings of the same feature, rather than waiting for BLV users to request alternatives, would scaffold a verification strategy that our participants discovered through necessity~\cite{bucinca_cognitive_forcing_2021}.

Interlocutors construct shared knowledge through coordinated speech acts~\cite{clark1996using}, and in extended conversation, this shared knowledge stabilizes into a reference vocabulary through lexical entrainment~\cite{brennan1996lexical, garrod1987saying}. Our findings mirror this pattern: detailed geometric descriptions served early in exploration, with shared labels stabilizing into reference points used throughout the session. The challenge for AI description systems is balancing the expressivity needed to build initial comprehension with the terminological consistency required to sustain it through the transition from use to editing.

An \textit{expression failure} occurs when a description contains accurate and complete information about a segment or structural relationship but presents it in a reference frame that does not map onto the BLV user's existing spatial knowledge, producing a misconception despite the description being factually correct. Expression failures are the most difficult to detect through sighted-expert evaluation because the information is present and accurate; the failure lies in the translation between the frame used and the mental model the user can construct from it. It seems unlikely that any method for assessing a description system that does not directly measure BLV participants' comprehension will be able to distinguish between accuracy, completeness, or expression failures---sighted evaluators are not sufficient proxies.

\subsection{Navigation: Scaffolding Model Exploration}
\label{sec:navigation}

Navigating a complete, expressive description of a 3D model poses challenges that cannot be resolved solely by improving the quality of the description. Prior work has demonstrated that the length of text descriptions scales with object complexity~\cite{Wong2022May}. As M5 noted, verbal descriptions of 3D models require substantially more time to convey what sighted access provides in seconds. Paradoxically, descriptions were simultaneously too long and too short due to the shift from parallel visual to sequential verbal information.

Breaking descriptions into conversational, on-demand snippets mitigates the length problem but introduces a distinct navigation challenge. When details are provided only upon request, BLV users must know what to ask to receive relevant information. Our participants consistently struggled to formulate useful questions, even when they recognized their understanding was incomplete. This aligns with Wu and Dunning's~\cite{wu_dunning_hypocognition_2018} phenomenon of \textit{hypocognition}: a lack of conceptual vocabulary to represent an idea or recognize its absence, leading users to assume that features omitted from the description must not exist (e.g., M1's assumption that the dragon had no claws). As a result, exploration of the model is heavily biased toward features described earliest in the interaction, leaving other aspects of the model underexplored and presumed to be nonexistent. BLV participants consistently devoted more attention to features described in the initial overview because these early interactions created a scaffold for subsequent inquiry. LLM-based systems may exacerbate this by producing fluent, confident-sounding descriptions that encourage passive acceptance over active interrogation~\cite{chromik_illusion_explanatory_depth_2021, bucinca_cognitive_forcing_2021}.

A \textit{navigation failure} occurs when relevant information exists within a description, but the BLV user does not reach it. Navigation failures can occur because details are delivered only after an inquiry stalls, hypocognition prevents the user from formulating the question needed to access them, or the initial description did not provide sufficient scaffolding for further exploration. Navigation failures are particularly deceptive: unlike accuracy or completeness failures, the information needed to cross the use-to-editing gap is present in the system but practically inaccessible to the user. A sighted evaluator reviewing the description would find nothing missing; only a BLV user navigating it under task pressure would encounter the failure.

The three criteria above suggest that natural language descriptions of 3D models are not a single-quality resource; they are a tool whose utility varies depending on what BLV makers need to do with a model. We organize this variation around a central distinction: the difference between \textit{using} a model, which requires sufficient comprehension to form a general impression of the model's structure, and \textit{editing} a model, which requires comprehension sufficient to specify changes to it. The distance between these two levels of understanding is the \textit{use-to-editing gap}.

\begin{strip}
\section{A Framework for Evaluating Visual-Description~Systems}\label{sec:implications}

\begin{center}
\captionof{table}{BLV participants self-rated confidence and a sighted~researcher's~assessment of the clay replicas.}\label{tab:blind_results}
\vspace{-10pt}
\begin{tabular}{@{}lcccccc@{}}
\multicolumn{1}{c}{\multirow{2}{*}{\textbf{ID}}} & \multicolumn{3}{c}{\textbf{First Model Comprehension}} & \multicolumn{3}{c}{\textbf{Second Model Comprehension}} \\
\multicolumn{1}{c}{} &
  \multicolumn{1}{c}{\textbf{Confidence}} &
  \multicolumn{1}{c}{\textbf{\begin{tabular}[c]{@{}c@{}}Structural\end{tabular}}} &
  \multicolumn{1}{c}{\textbf{\begin{tabular}[c]{@{}c@{}}Segment\end{tabular}}} &
  \multicolumn{1}{c}{\textbf{Confidence}} &
  \multicolumn{1}{c}{\textbf{\begin{tabular}[c]{@{}c@{}}Structural\end{tabular}}} &
  \multicolumn{1}{c}{\textbf{\begin{tabular}[c]{@{}c@{}}Segment\end{tabular}}} \\ \midrule
\textbf{M1}  & 6 & 6 & 4 & 4 & 1 & 1 \\
\textbf{M2}  & 3 & 3 & 1 & 2 & 6 & 2 \\
\textbf{M3}  & 6 & 3 & 6 & 4 & 5 & 4 \\
\textbf{M4}  & 6 & 4 & 6 & 5 & 7 & 7 \\
\textbf{M5}  & 4 & 6 & 3 & 5 & 4 & 4 \\
\textbf{M6}  & 7 & 1 & 1 & 7 & 1 & 1 \\
\textbf{M7}  & 4 & 2 & 1 & 5 & 3 & 2 \\
\textbf{M8}  & 6 & 4 & 3 & 3 & 2 & 1 \\
\textbf{M9}  & 7 & 1 & 1 & 7 & 1 & 1 \\
\textbf{M10} & 5 & 6 & 3 & NA & NA & NA \\ \bottomrule
\end{tabular}
\end{center}

\begin{center}
\begin{minipage}[b]{.18\linewidth}
    \centering
    \includegraphics[height=85pt, alt={A model of a hip and pelvis bone connected to the top of the femur. The hip and femur bones are printed with organic hole structures dispersed throughout the model. The interior of the bones is hollow. The hip bone starts with a wide, flaring area and then slopes into a wide, inward-pointing area. The top of the femur is a rough ball-like shape that slots into the lower hip bone.}]{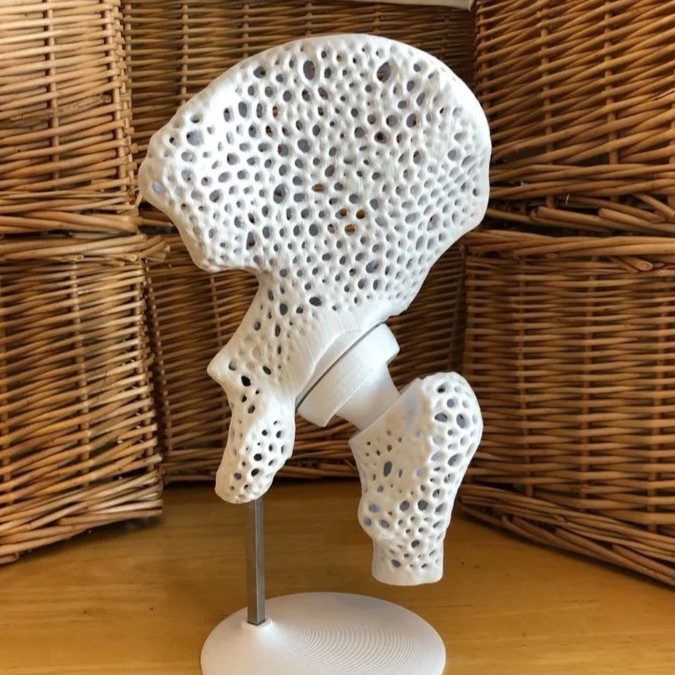}
    \Description[A photograph from Thingiverse of the hip-pelvis model]{A model of a hip and pelvis bone connected to the top of the femur. The hip and femur bones are printed with organic hole structures dispersed throughout the model. The interior of the bones is hollow. The hip bone starts with a wide, flaring area and then slopes into a wide, inward-pointing area. The top of the femur is a rough ball-like shape that slots into the lower hip bone.}
\end{minipage}%
\begin{minipage}[b]{.18\linewidth}
    \centering
    \includegraphics[height=85pt, alt={The pelvis segment is a medium-sized ball with an extension out to cup around a similarly sized ball at the top of the femur. The femur segment has a long rod extending out from the bottom of the ball, which forms the hip joint.}]{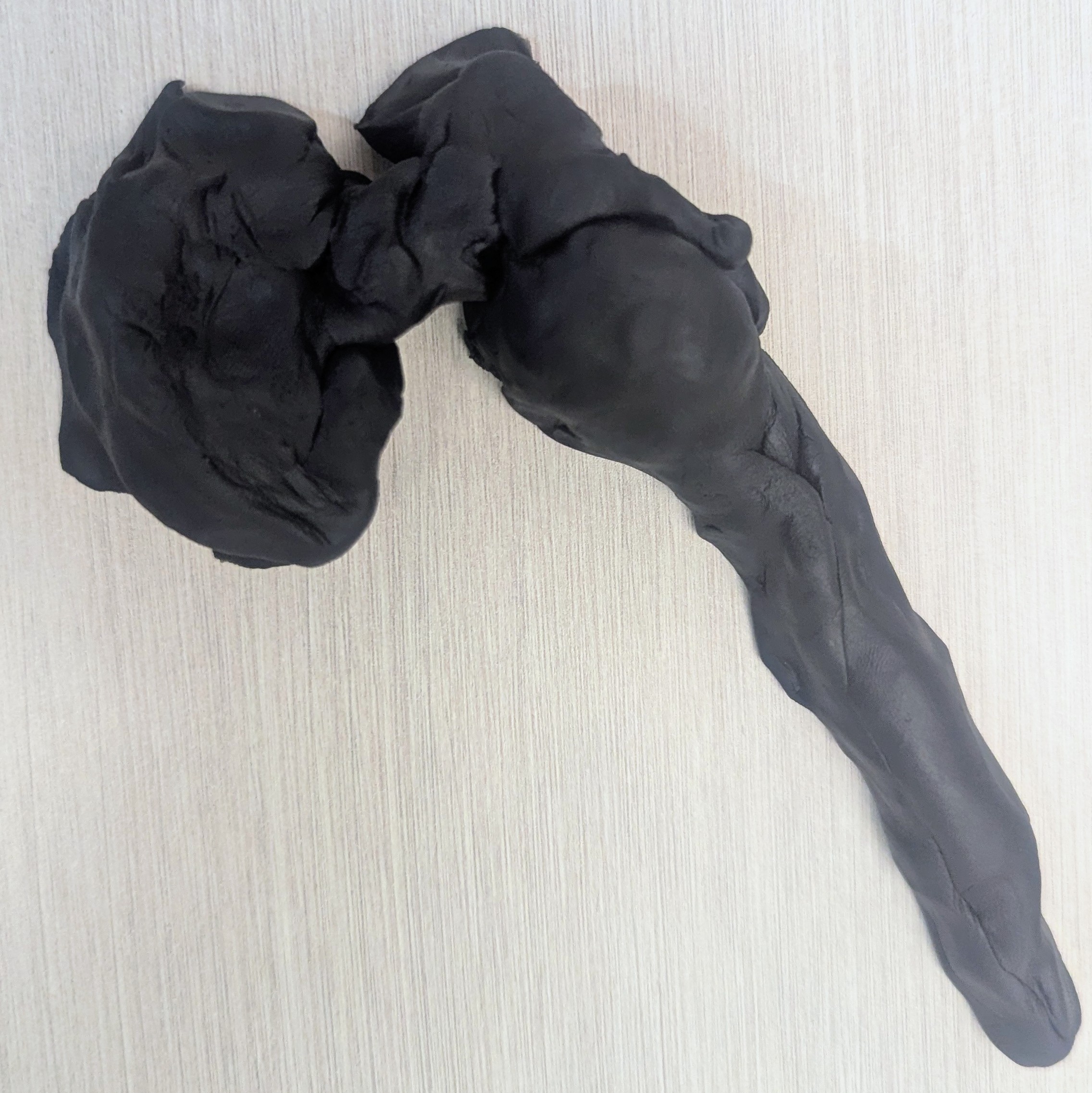}
    \Description[A photograph of the black clay model of the hip-pelvis made by M2]{The pelvis segment is a medium-sized ball with an extension out to cup around a similarly sized ball at the top of the femur. The femur segment has a long rod extending out from the bottom of the ball, which forms the hip joint.}
\end{minipage}%
\begin{minipage}[b]{.36\linewidth}
    \centering
    \includegraphics[height=85pt, alt={A photograph of the Supreme Court Building model, capturing its iconic neoclassical architectural design. The model includes detailed features of the building's structure. The building has a rectangular footprint, with a prominent central section flanked by symmetrical wings. The central section features a grand entrance accessible via a wide staircase. The stairs lead up to a portico with a row of columns supporting a triangular pediment. The wings on either side are rectangular, with flat roofs and uniformly spaced windows.}]{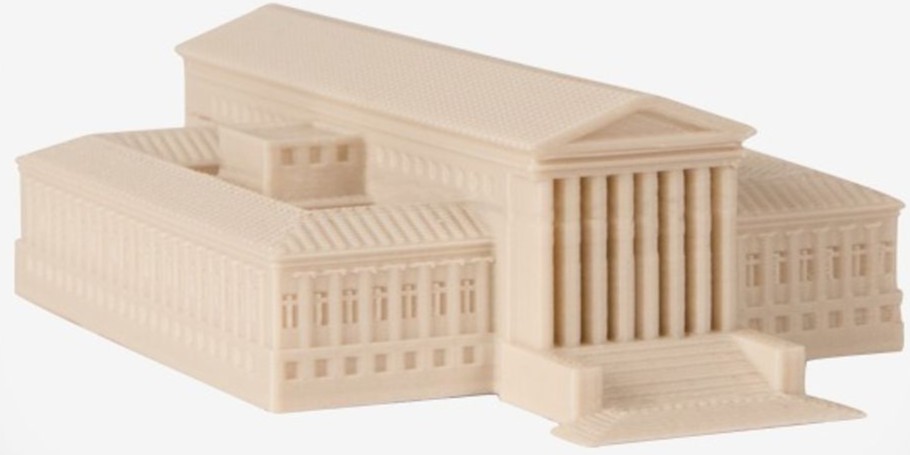}
    \Description[A photograph of the Thingiverse Supreme Court model]{A photograph of the Supreme Court Building model, capturing its iconic neoclassical architectural design. The model includes detailed features of the building's structure. The building has a rectangular footprint, with a prominent central section flanked by symmetrical wings. The central section features a grand entrance accessible via a wide staircase. The stairs lead up to a portico with a row of columns supporting a triangular pediment. The wings on either side are rectangular, with flat roofs and uniformly spaced windows.}
\end{minipage}%
\begin{minipage}[b]{.25\linewidth}
    \centering
    \includegraphics[height=85pt, alt={The building front includes a series of stairs with 2 columns on top of the stairs, framing a small doorway. The building is a wide rectangular structure with a small pointed roof over the columns of the stairs.}]{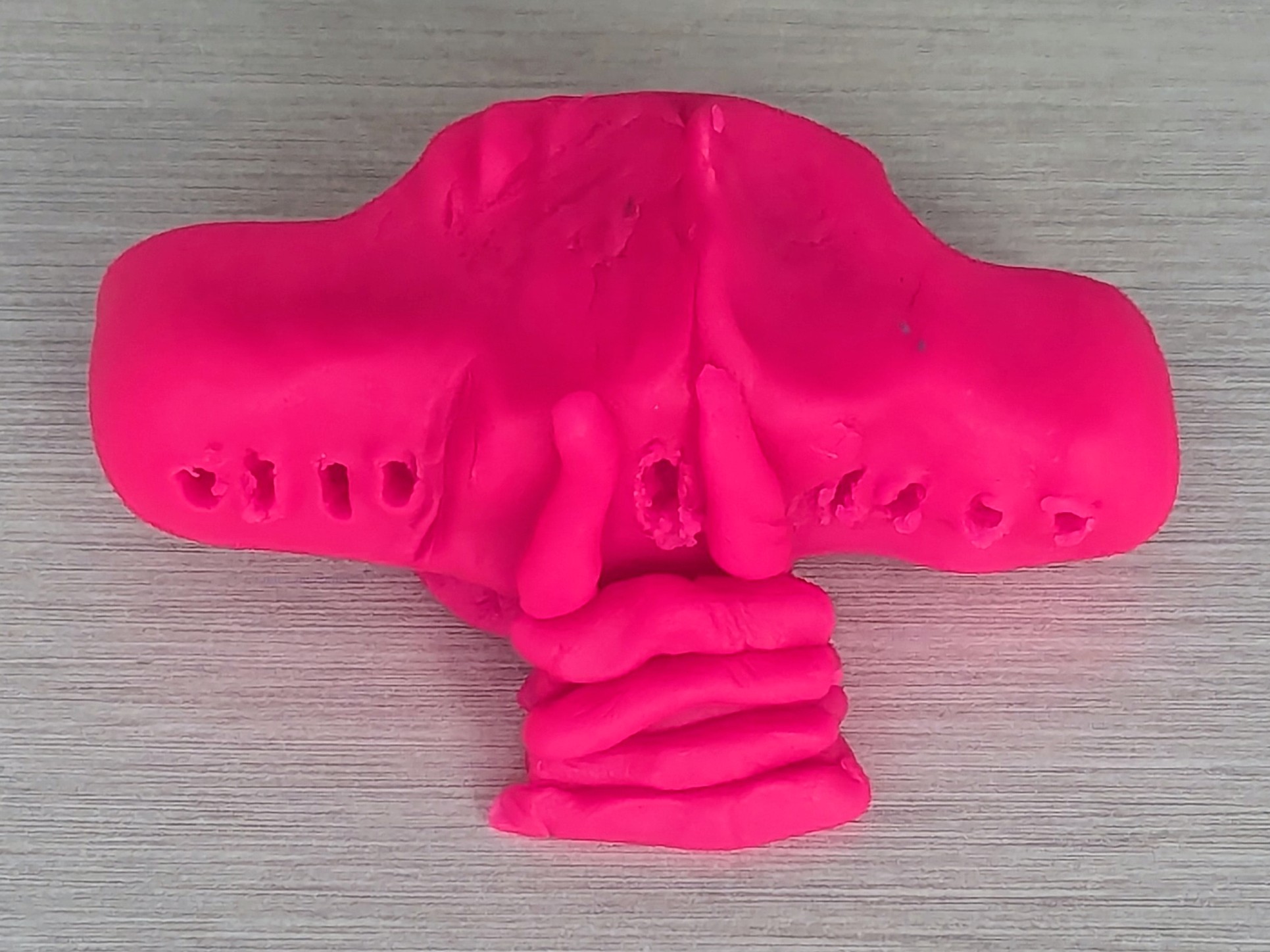}
    \Description[A photograph of the pink clay model of the Supreme Court building made by M5]{The building front includes a series of stairs with 2 columns on top of the stairs, framing a small doorway. The building is a wide rectangular structure with a small pointed roof over the columns of the stairs.}
\end{minipage}
\hfill
\captionof{figure}{Examples of printed models from Thingiverse and the clay replicas created by M2 (left) and M5 (right).}\label{fig:example_evaluations}
\end{center}
\end{strip}

We propose a methodology for evaluating whether a description system enables BLV users to cross this gap. The test involves two phases: first, assessing BLV participants' spatial mental models of a 3D model, and second, assessing whether that comprehension supports specification of a modification. The second test is falsifiable: if modification is impaired by low model comprehension, then the generated description is insufficient for systems in which BLV users edit models. Analysis of the first and second tests will reveal different sources of description failures, which can inform the design of the final system.

\subsection{Phase 1: Comprehension Assessment}

After BLV participants explore a model through a visual description, researchers assess their comprehension of the model. A comprehensive spatial mental model will accurately represent each \textit{segment} of the 3D model and their structural relationships. We adopt the distinction between individual segments and their structural relationships from Faruqi et al.'s taxonomy of functional models~\cite{Faruqi_Style2Fab}. 

We present four criteria for rating model comprehension. To illustrate the rating process, \autoref{tab:blind_results} presents a comprehension evaluation for all BLV participants, and \autoref{fig:example_evaluations} compares the original Thingiverse models with clay replicas created by M2 and M5 to contextualize our evaluation criteria.

\subsubsection{Segment Comprehension}

Segments are salient independent sections of a model (e.g., limbs of an animal figurine, regions that serve a specific functional purpose). Individual segments are often the site of editing tasks, such as changing textures, colors, or size parameters.

\paragraph{Segment Accuracy} is assessed by identifying errors in the participant's mental model of each segment. Does the replica include features of a segment that do not exist? Are features misinterpreted to a degree that would misdirect an edit? For example, M2's low segment-level comprehension of the hip model is partly attributable to a pelvis segment that did not match the original model's shape.

\paragraph{Segment Completeness} is assessed by identifying features present in the original segment but absent from the participant's mental model. Regardless of whether a feature was explicitly described, the participant should be aware of all salient features of the segment. M2 also showed low segment completeness by omitting its hole-filled organic texture on the pelvis. M5 showed a less severe completeness gap by reproducing only two columns at the Supreme Court portico rather than the full row.

\subsubsection{Structural Comprehension}

Structural relationships define how segments fit together to form the whole model. Many of the editing goals expressed by our BLV participants involved structural changes: altering the relative positions of segments (e.g., spacing between fingers, M7) or adding new segments in specific spatial relationships to existing ones (e.g., connecting gears, M4).

\paragraph{Structural Accuracy} is assessed by identifying errors in how the participant arranged segments. No extra segments absent from the original should be present. Each segment should connect correctly to its neighbors and maintain the original proportions. M2's structural score, while higher than their segment score, reflects an inaccurate proportion of the pelvis segment relative to the femur. M5 merged the wings of the Supreme Court building with the central section, conflating two distinct structural segments.

\paragraph{Structural Completeness} is assessed by identifying missing segments or spatial relationships. Even if certain relationships were not described, the participant should understand how all segments compose the whole. M5, for example, omitted the walls bounding the sides of the stairwell.

\subsubsection{Conducting Comprehension Assessments}

In our study, a physical reproduction task using clay replicas provided a comprehensive expression of participants' spatial mental models. Accuracy and completeness errors were visible in the artifacts, and post-hoc interviews enabled diagnosis of expression and navigation failures by probing which information participants received, how they interpreted it, and how their understanding shifted over the course of the interaction. Other study contexts may call for different assessment methods, such as structured verbal protocols or survey instruments, depending on the deployment setting and participant population.

\subsection{Phase 2: Editing Specification Assessment}

Once participants demonstrate their comprehension of a model, researchers test whether that comprehension supports crossing the use-to-editing gap. Participants are given a modification goal and an editing method suited to the deployment context (e.g., our structured prompting protocol, a programming-based tool~\cite{zhang_a11yshape_2025}, or a prototype CAD interface). Their editing specifications are then assessed on two criteria:

\begin{itemize}
    \item \textit{Segment Specificity}: Does the participant correctly identify the segment to be modified and specify the change with sufficient precision to execute without clarification?

    \item \textit{Structural Dependency Awareness}: Does the specification account for how the modification interacts with adjacent segments, or does it inadvertently create contradictions?
\end{itemize}

Researchers should trace Phase 2 failures back to Phase 1 comprehension errors. A participant who fails segment specificity after passing Phase 1 segment comprehension likely faces a navigation failure: the relevant information was present in the description but was not accessible at the moment of editing. A participant who fails structural dependency awareness likely carries an accuracy or completeness failure forward from Phase 1: a misrepresented or absent structural relationship produced a specification that does not account for how segments interact.

\subsection{Designing Systems after Failure Diagnosis}

Diagnosing which type of failure a system produces directs developers towards different approaches to improving the system. Improving description quality as an undifferentiated target is unlikely to close the use-to-editing gap; the appropriate intervention depends on where the test reveals descriptions are falling short.

\subsubsection{Addressing Accuracy and Completeness Failures}

Accuracy and completeness failures are primarily a function of the 3D model representation a system uses. A system that generates descriptions from multi-view rendered images can produce accurate overviews of external geometry, but cannot convey internal structure, occluded features, or precise contours. At best, the resulting descriptions will be incomplete; at worst, AI systems will hallucinate to fill the representational gap~\cite{hullucination_mitigation}. 

Addressing these failures requires expanding what a system can access: non-Euclidean representations such as 3D meshes provide greater geometric detail but impose significant computational costs~\cite{survey_3D_Data_Rep_for_DL}, while rich feature sets that explicitly expose structural properties may reduce the need to infer geometry from insufficient input~\cite{RAG}. Where representational limits cannot be remediated, they should be disclosed to BLV users rather than concealed, consistent with seamful XAI design principles~\cite{Ehsan_Seamful_XAI, inman_seams}: a system that articulates what it can and cannot access from its input representation allows BLV users to distinguish an AI's silence from a model's absence.

\subsubsection{Addressing Expression Failures}

Expression failures call for interventions at two levels. At the interface level, systems can make reference frame switching explicit and accessible. A11yShape~\cite{zhang_a11yshape_2025} offers one example: by synchronizing AI-generated descriptions with a hierarchical model structure and source code, it grounds descriptions in multiple simultaneous representations, allowing BLV users to navigate between a semantic hierarchy, a textual description, and a code-level specification. This cross-representational anchoring reduces the cost of frame-switching. At the description generation level, Levinson's taxonomy of reference frames~\cite{levinson_space_language_2003} provides a design vocabulary: systems that generate descriptions in multiple frames for the same feature, and that allow users to request a specific frame, scaffold the verification-by-consistency strategy our participants developed spontaneously. Expression failures rooted in a mismatch between a BLV user's prior spatial knowledge and the frames a description assumes cannot be fully resolved at the generation level; they also require system support to help users anchor new descriptions to their existing knowledge.

\subsubsection{Addressing Navigation Failures}

Navigation failures point to interventions in how descriptions are structured and delivered relative to a user's task. A description system that presents a fixed, sequential overview cannot tailor its content to what a specific BLV user needs to know to execute a given edit. Two complementary approaches address this. First, systems can incorporate contextual information beyond the 3D model itself: the description posted by the model creator in a repository, the intended purpose of the model, and the known modification space (i.e., which parameters are designed to be adjustable) all provide scaffolding that reduces the burden on the user to discover relevant information through open-ended inquiry. Second, systems can tune descriptions to a series of posed questions rather than generating a fixed overview: a system that knows a BLV user wants to replace text on a sign element can lead with a description of that element, its location, and its relationship to adjacent segments, providing the needed information without requiring the user to navigate to it. This task-responsive design directly addresses navigation failures by expanding the scaffold for inquiry beyond what an initial overview alone provides.

\newpage
\section{Limitations}

We acknowledge that real-world experience with 3D modeling and related technical disciplines is relatively rare in the BLV community, partly because these tools are difficult to access. Despite efforts to recruit experienced BLV modelers, we were able to recruit only two BLV participants with prior experience with 3D modeling tools. As a result, the main study scripts were developed from descriptive strategies elicited by the blind first author, whose particular background in engineering-oriented 3D modeling may not generalize to BLV modelers with different domains of expertise or prior experience. Sighted assistants may also develop alternative descriptive approaches when working with modelers of different backgrounds. These approaches did not surface in our pilot and did not inform script design. This distinction between novice modelers and experts is not unique to BLV users; makers often fall along a spectrum from novice to expert-amateur to professional, which confounds the design of general-purpose CAD tools. Indeed, an ideal approach may be to create a variety of accessible modeling tools tailored to different expertise levels, as the current market of visual CAD tools accommodates a range of experience levels (e.g.,~\cite{tinkercad, fusion}).

Our simulation of AI agents is not fully representative of a real system. This trade-off was deliberate: scripting all descriptions allowed us to test the ceiling of what language-based descriptions can support, independent of any particular AI system's reliability. However, when BLV participants asked questions without relevant information in the script, sighted researchers consistently declined to answer---whereas an AI agent would likely hallucinate a response that could mislead the BLV user. An AI agent could also be built with 3D model embeddings that include far more information than we included in our scripts. Yet embedding all seemingly relevant detail into a 3D model representation remains a significant open challenge~\cite{survey_3D_Data_Rep_for_DL}. As a result, we may not have fully accounted for either the capabilities of future AI agents or the ways BLV users respond to inaccurate or hallucinated information.

Our clay replica method provides a rich qualitative window into participants' spatial mental models, but it is an imperfect proxy for comprehension. We cannot expect any modeler--blind or sighted--to sculpt an accurate replica, and replica quality depends on motor skills and prior clay experience as much as on model understanding. Our comprehension assessments were conducted by a single-sighted researcher without inter-rater reliability checks, and our 3D models were too diverse to support systematic comparison across participants. These constraints motivated the more structured evaluation protocol proposed in the framework section: the two-phase comprehension and editing specification assessment provides a methodology that future work can apply more rigorously, with validated criteria, multiple raters, and participant pools large enough to permit comparison across description conditions.

\newpage
\section{Conclusion}

Through an autoethnographic pilot with twelve sighted participants and a Wizard-of-Oz main study with ten BLV participants, we found a \textit{use-to-editing gap} between the level of comprehension natural language descriptions of 3D models can produce and the spatial understanding needed to meaningfully edit 3D models. This gap produces four failure types that evaluation approaches relying solely on sighted expert assessment cannot detect. Accuracy and completeness failures are visible in artifacts like clay replicas, but expression failures (correct information in a misaligned reference frame) and navigation failures (relevant information present but never reached) are invisible to any evaluator who does not directly observe a BLV user navigating a description under task pressure.

To address this, we propose a two-phase evaluation framework. Phase 1 assesses BLV participants' spatial mental models; Phase 2 assesses whether that comprehension supports editing specification. Tracing Phase 2 failures back to Phase 1 diagnoses identifies which failure type is operative and directs developers toward targeted interventions. We offer this framework as an evaluation methodology that future work can refine and apply to assess whether description-based systems genuinely support BLV modelers in their role as creators.

\begin{acks}
We thank our participants for their time and effort towards this work. This research was funded by the National Science Foundation (NSF: CISE 2327137).
\end{acks}

\bibliographystyle{ACM-Reference-Format}
\bibliography{references}

\newpage
\onecolumn
\appendix
\section{Sighted-Assistant Pilot Study Participants}\label{app:sighted_participants}
\begin{table*}[h]
\centering
\caption{Summary demographics for sighted participants, referenced in the paper by the letter `A' and an identifying number.}
\label{tab:assistants}
\vspace{-10pt}
\begin{tabular}{@{}lllll@{}}
\toprule
\textbf{Identifier} & \textbf{Gender} & \textbf{Is Domain Expert} & \textbf{Has Sighted-Assistant Experience} & \textbf{Models} \\ \midrule
A1  & Female & No                     & No  & 1,2   \\
A2  & Female & No                     & No  & 3,4   \\
A3  & Male   & Yes      & No  & 5,6   \\
A4  & Female & Yes      & Yes & 7,8   \\
A5  & Male   & Yes      & Yes & 9,10  \\
A6  & Female & No                     & Yes & 11,12 \\
A7  & Female & No & Yes & 13,14 \\
A8  & Female & No                     & Yes & 15,16 \\
A9  & Male   & Yes      & No  & 17,18 \\
A10 & Female & Yes   & No  & 19,20 \\
A11 & Female & Yes   & Yes & 21,22 \\
A12 & Female & No                     & No  & 23,24 \\ \bottomrule
\end{tabular}
\end{table*}

\section{Likert Scale Assessments from Sighted-Assistant Study}
\begin{table*}[ht!]
\centering
\caption{Results of Likert Scale assessment of blind first author's comprehension of the models demonstrated by their self-rated confidence and a sighted researcher's assessment of the clay replicas.}
\label{tab:sighted_results}
\vspace{-10pt}
\begin{tabular}{@{}lcccccc@{}}
\multicolumn{1}{c}{\multirow{2}{*}{\textbf{ID}}} & \multicolumn{3}{c}{\textbf{First Model Comprehension}} & \multicolumn{3}{c}{\textbf{Second Model Comprehension}} \\
\multicolumn{1}{c}{} &
  \multicolumn{1}{c}{\textbf{Confidence}} &
  \multicolumn{1}{c}{\textbf{\begin{tabular}[c]{@{}c@{}}Structural\end{tabular}}} &
  \multicolumn{1}{c}{\textbf{\begin{tabular}[c]{@{}c@{}}Segment\end{tabular}}} &
  \multicolumn{1}{c}{\textbf{Confidence}} &
  \multicolumn{1}{c}{\textbf{\begin{tabular}[c]{@{}c@{}}Structural\end{tabular}}} &
  \multicolumn{1}{c}{\textbf{\begin{tabular}[c]{@{}c@{}}Segment\end{tabular}}} \\ \midrule
\textbf{A1}                                      & 6            & 6           & 5           & 5            & 7            & 3           \\
\textbf{A2}                                      & 5            & 3           & 3           & 6            & 5            & 5           \\
\textbf{A3}                                      & 6            & 6           & 6           & 5            & 4            & 6           \\
\textbf{A4}                                      & 3            & 5           & 2           & 7            & 7            & 2           \\
\textbf{A5}                                      & 3            & 7           & 6           & 3            & 1            & 1           \\
\textbf{A6}                                      & 3            & 4           & 4           & 7            & 7            & 5           \\
\textbf{A7}                                      & 7            & 7           & 5           & 5            & 3            & 3           \\
\textbf{A8}                                      & 3            & 1           & 1           & 6            & 6            & 5           \\
\textbf{A9}                                      & 6            & 7           & 7           & 3            & 6            & 7           \\
\textbf{A10}                                     & 4            & 5           & 4           & 4            & 5            & 5           \\
\textbf{A11}                                     & 5            & 5           & 4           & 7            & 7            & 4           \\
\textbf{A12}                                     & 5            & 6           & 5           & 7            & 7            & 4           \\ \bottomrule
\end{tabular}
\end{table*}

\section{Thingiverse 3D Models Used in Study}\label{app:models}
\begin{table*}[t]
\centering
\caption{A summary of the 3D models collected for these studies. Models are presented in the order used in the studies. The~Thingiverse ID numbers can be used to find the source models with URLs structured as ``www.thingiverse.com/thing:\#\#''. }
\label{tab:models}
\vspace{-10pt}
\begin{tabular}{@{}cccc@{}}
\toprule
\textbf{Model} & \textbf{Brief Description} & \textbf{Thing ID} & \textbf{Category} \\ \midrule
1  & Dragon                   & 6635990 & Figurine             \\
2  & Custom QR Code Menu Card & 6640744 & Real-World Object    \\
3  & Charger Mount            & 6636110 & Functional Component \\
4  & Hip and Pelvis Model     & 4875641 & Anatomical           \\
5  & Self Balancing Bird      & 6652740 & Figurine             \\
6  & Statue of Liberty        & 65869   & Real World Object    \\
7  & Gear Bearing             & 53451   & Functional Component \\
8  & Low-Poly Skull           & 906562  & Anatomical           \\
9  & Vulpix Pokemon           & 3455495 & Figurine             \\
10 & Supreme Court Building   & 464195  & Real World Object    \\
11 & Anatomical Heart         & 932606  & Anatomical           \\
12 & Capybara                 & 182089  & Figurine             \\
13 & Lizard                   & 6650989 & Figurine             \\
14 & Paris Building           & 30649   & Real World Object    \\
15 & Flexy-Hand               & 242639  & Anatomical           \\
16 & 12V Power Supply Cover   & 40377   & Functional Component \\
17 & Low-Poly Cat             & 908621  & Figurine             \\
18 & Seattle Space Needle     & 567695  & Real-World Object    \\
19 & Anatomic Human Foot      & 22628   & Anatomical           \\
20 & Universal Joint          & 264017  & Functional Component \\
21 & Bolt                     & 193647  & Functional Component \\
22 & Forbidden Watchtower     & 2574499 & Real World Object    \\
23 & Spinal Cord              & 6675485 & Anatomical           \\
24 & RC Chassis               & 3576552 & Functional Component \\
\bottomrule
\end{tabular}
\end{table*}

\section{3D Model Descriptions} \label{app:3D_Model_Descriptions}

\paragraph{{1. Dragon}}
An Asian dragon figurine with a long snake-like body made of many articulating segments. The figurine starts with a head with two horns connected to many segments. Two segments have two legs on either side of the body attached. The figurine ends with a short tail segment. The dragon is lying down with its legs at right angles as if crouching. 

\paragraph{{2. Custom QR Code Menu Card}}
A menu card stands that holds a card with an embossed customizable symbol, such as a sign that says "Check-In," an email symbol, or a WiFi symbol. Above the symbol is space to emboss a square QR code. The stand has a rectangular base with a rectangular hole extruded at a slight angle. This hole is wide enough to hold the short end of the menu card. The rectangular hole on the back side of the card extrudes a rounded backing that can hold the card at an angle. 

\paragraph{{3. Charger Mount}}
A USB charger mount for an Arduino's TA4045 Lithium battery charger module. The model has a rounded rectangular base with a hole to put the male end of a USB cable. To the left and right of the USB cable hole are two screw holes with rounded screw head sinks. Perpendicular to the base is a thin rectangular backing the size of the PCB. The edge of the backing is an embossed frame. At the top of the backing are two pairs of evenly spaced holes. At the bottom of the backing, near the base, there are two holes on either side. 

\paragraph{{4. Hip and Pelvis Model}}
A model of a hip and pelvis bone connected to the top of the femur. The hip and femur bones are printed with organic hole structures dispersed throughout the model. The interior of the bones is hollow. The hip bone starts with a wide flaring area and then slopes into an inward-pointed wide area. The top of the femur is a rough ball-like shape that slots into the lower hip bone. 

\paragraph{{5. Self Balancing Bird}}
Represents a self-balancing bird, intricately designed to balance on its beak. The bird exhibits a lifelike and dynamic posture, with wings outstretched as if in mid-flight. The bird's body is sleek and aerodynamic, featuring a streamlined shape that suggests it is soaring through the air. Its head is small and rounded, with a prominent beak that extends forward. The beak, slightly curved and pointed, serves as the balance point for the entire model. This design allows the bird to balance precisely on the tip of its beak, creating a captivating visual effect. The wings of the bird are fully extended and curved downward, forming a semicircular shape. Each wing is adorned with detailed feathers, which are evenly spaced and slightly raised to add texture and a realistic appearance. The feathers taper gracefully towards the tips, enhancing the sense of motion and flight. The tail feathers are fanned out and extended, complementing the curvature of the wings and adding to the overall balance of the design. These tail feathers are also detailed with individual feather patterns, contributing to the bird's realistic look. The underside of the bird, from its body to its wings and tail, is smooth, allowing for a clean and polished appearance. The overall symmetry and precise detailing of the feathers create an aesthetically pleasing and balanced structure. 

\paragraph{{6. Statue of Liberty}}
A detailed representation of the Statue of Liberty, including its pedestal base. The statue stands tall and majestic, capturing the essence of this famous monument. The statue itself depicts Lady Liberty, with her right arm raised high, holding the torch. The arm is extended upwards, with the torch's flame prominently positioned at the top. Lady Liberty's left-hand holds a tablet close to her body. The statue's robe flows naturally around her, with folds and drapes that add realism and movement to the model. Her crown features seven spikes, each spike evenly spaced and protruding from her head. The pedestal base is a square structure with multiple levels, providing a sturdy foundation for the statue. The base is detailed with architectural features such as columns and reliefs, which add to the model's authenticity. The bottom level of the pedestal is the widest, tapering slightly as it rises, with each tier featuring decorative elements that mimic the actual monument. Overall, the proportions of the model are accurate, with the statue's height and the base's dimensions balanced to reflect the real-life structure. The fine details of Lady Liberty's attire, the torch, the tablet, and the base's architectural design make this model a precise and impressive representation of the Statue of Liberty. 

\paragraph{{7. Gear Bearing}}
Represents a gear bearing, showcasing an intricate design of interlocking gears within a circular frame. The bearing consists of multiple gears arranged in a precise configuration to allow smooth rotational movement. The outer frame of the gear bearing is circular, providing a sturdy boundary that houses the internal gears. The surface of the frame is smooth, with a slight lip around the edge to keep the gears securely in place. The interior of the frame features a series of evenly spaced notches that match the gears' teeth, ensuring proper alignment and engagement. Inside the frame, there are several gears with a star-like appearance, each with multiple pointed teeth. These gears are arranged in a symmetrical pattern around a central gear. The teeth of each gear are designed to interlock seamlessly with the adjacent gears, allowing them to rotate together smoothly. The gears have a textured surface, adding to the visual complexity and providing better grip during rotation. At the center of the gear bearing is a larger, central gear that serves as the main hub. This gear is connected to the surrounding gears, ensuring that when it rotates, all other gears move in unison. The central gear has a hole in the middle, likely for mounting or attaching to another component. 

\paragraph{{8. Low-Poly Skull}}
Features a low poly skull, giving it a distinct geometric appearance. The skull's surface is composed of numerous flat, triangular facets, creating a faceted and angular look typical of low-poly designs. The top of the skull is a rounded dome shape but with flat surfaces instead of smooth curves. The front of the skull features prominent eye sockets, which are large, deep, and diamond-shaped, positioned symmetrically. Below the eye sockets, the nasal cavity is a small, inverted triangle that adds to the overall polygonal theme. The jawline is angular, with sharp edges defining the chin and jaw structure. The mouth area is simplified, consisting of a hollow space that hints at the teeth without intricate detailing. The cheekbones are also pronounced, with angular surfaces that extend outward. Overall, this low poly skull model captures the essential form of a human skull but does so using flat surfaces and sharp angles, creating a stylized and modern interpretation. 

\paragraph{{9. Vulpix Pokemon}}
Represents Vulpix, a popular Pokémon character, depicted in a stylized and low-poly design. Vulpix has a distinctively fox-like appearance with several unique features that are captured in this model. The body of Vulpix is compact and pear-shaped, sitting in a natural, relaxed pose with all four legs touching the ground. Each leg is short and sturdy, ending in small, blocky paws that support the character's weight. The legs are slightly bent, giving the impression of a seated position. The head of Vulpix is large in proportion to its body, with a smooth, angular shape. The eyes are represented by indented oval shapes, and the nose and mouth are similarly simple geometric forms, all adding to the low-poly aesthetic. The ears are triangular and pointy, protruding from the top of the head, and angled slightly outward. One of Vulpix's most recognizable features is its large, bushy tail. This model showcases the tail as a series of sweeping, curled segments that extend from the back of the body. The tail is voluminous and fanned out, resembling a series of curled, overlapping petals or waves. Each segment is carefully angled to give the tail a dynamic, flowing appearance. 

\paragraph{{10. Supreme Court Building}}
Represents the Supreme Court Building, capturing its iconic neoclassical architectural design. The model includes detailed features of the building’s structure, providing a miniature yet accurate representation. The building has a rectangular footprint, with a prominent central section flanked by symmetrical wings on either side. The central section features a grand entrance, accessible via a wide staircase. The stairs lead up to a portico with a row of columns supporting a triangular pediment. The columns are evenly spaced and cylindrical, adding to the grandeur of the entrance. The roof of the central section is flat and slightly pitched, with a rectangular block extending upwards in the middle, representing the upper part of the courtroom. The wings on either side of the central section are also rectangular, with flat roofs and uniformly spaced windows. Each window is represented by small indentations along the length of the walls, creating a consistent and orderly appearance. The base of the model is flat, ensuring stability when placed on a surface. The sides and back of the building mirror the front's design, with consistent architectural elements and window placement. 

\paragraph{{11. Anatomical Heart}}
Accurately represent the anatomical features of the heart. The model captures the complex structure of the heart, including its major chambers and associated blood vessels. It is an asymmetric, somewhat conical organ with rounded, bulging surfaces. At the top of the heart are several large blood vessels, including the aorta and pulmonary arteries. These vessels are cylindrical and protrude from the main body of the heart, each with smooth, rounded openings. The vessels vary in size, with the aorta being the largest. The surface of the heart is detailed with various grooves and ridges that correspond to the heart's muscular structure and blood flow pathways. The left and right ventricles form the bulk of the heart's lower portion, with the left ventricle being slightly larger and more rounded than the right. The right and left atria sit above the ventricles, appearing as smaller, slightly rounded chambers. The overall texture of the heart is smooth but with enough detailing to show the separation between different parts, such as the atria, ventricles, and the septum that divides the left and right sides of the heart. There are also subtle indications of the heart valves, though they are not deeply indented. The heart's backside features additional vascular structures, including veins that return blood to the heart. These are less prominent than the arteries but are still clearly defined. In terms of proportions, the heart is roughly as tall as it is wide, with the major blood vessels adding height and dimension to the upper part.

\paragraph{{12. Capybara}}
Realistic and detailed design. The capybara is shown standing on all four legs, with a sturdy and robust body that reflects its real-life counterpart's physical characteristics. The body of the capybara is elongated and barrel-shaped, with a slight arch along the back. The surface of the model is textured with a fine, fur-like pattern that adds realism and depth to the figure. The legs are short and sturdy, with thick, cylindrical shapes that support the weight of the body. Each leg ends in a round, slightly flattened paw with distinct toes. The head of the capybara is large and blocky, with a broad, rounded snout that protrudes forward. The eyes are small and oval-shaped, positioned on the sides of the head, giving the capybara a gentle and calm expression. The ears are small and rounded, sitting just behind the eyes on top of the head. The nose is slightly upturned, and the mouth is a simple line that curves downward at the corners, adding to the overall docile appearance of the animal. The neck is short and thick, seamlessly connecting the head to the body. The tail of the capybara is very short and almost imperceptible, staying true to the actual animal's anatomy. 

\paragraph{{13. Lizard}}
A thin pencil-shaped Gecko whose body is composed of grooved scales, resembling the look of a spine. This is due to the fact that there are cuts between each scale as the body extends, and the end of the last scale is tipped\textbackslash{}rounded off. The middle scales get slightly wider between the sets of legs to portray the geckos belly. It lays flat on the floor and the feet stick out perpendicularly of the body. The front feet have three toes, while the rear ones have four each. While all the feet point sideways, the back legs protrude straight out and then bend 90 degrees towards the rear, making them parallel to the body, and the front feet come straight out and then bend 45 degrees away from the body. Its eyes are composed of two circles, one sticking out and one engraved in the face. It has a rounded nose. The body is about five times longer than the head. 

\paragraph{{14. Paris Building}}
Represents a classic Parisian building, characterized by its elegant architectural design and intricate details. The building features a rectangular footprint with a flat roof and multiple levels, capturing the essence of traditional Parisian architecture. The facade of the building is adorned with a series of tall, rectangular windows, evenly spaced across each level. These windows are framed with decorative moldings, adding a touch of sophistication and depth to the structure. Below each window, there are detailed ledges that extend slightly outward, enhancing the building's ornate appearance. The building's corners are accented with vertical columns that run from the base to the roof, providing structural emphasis and aesthetic balance. The base of the building features a series of larger windows and doors, indicating the presence of ground-floor entrances and possibly commercial spaces. Above the windows, the roofline is defined by a decorative cornice that wraps around the entire perimeter of the building. This cornice is detailed with intricate patterns and adds a crowning touch to the structure. The roof itself is flat and slightly raised in the center, creating a subtle but distinct profile. 

\paragraph{{15. Flexy-Hand}}
A detailed and functional model of a prosthetic hand. The Flexy-Hand is designed to mimic the natural movement and flexibility of a human hand. The hand is composed of multiple articulated segments, each representing a finger or part of the hand. The overall structure includes the palm, five fingers, and the wrist, with a design focused on replicating the joints and motion of a real hand. The palm is the central part of the hand, providing the base for finger attachment. It is slightly curved, resembling the natural shape of a human palm, and has a smooth surface. The back of the palm features attachment points for tendons or wires that facilitate movement. Each of the five fingers is composed of several segments (phalanges) connected by joints. The fingers include the thumb, positioned opposable to the other fingers, allowing for gripping. The thumb has two main segments and a rounded tip. The index, middle, ring, and pinky fingers each have three segments. The segments are cylindrical and connected by flexible joints, allowing for bending at the knuckles. The joints between the finger segments are designed to be flexible, enabling realistic movement. The wrist segment is attached to the base of the palm and is designed to connect the hand to the forearm or a prosthetic device. The wrist has a smooth, rounded shape and includes attachment points for cables or tendons. The Flexy-Hand's design allows for a wide range of motion, with each finger capable of bending at the joints. This articulation makes the hand suitable for grasping and manipulating objects, similar to a real hand. The attention to detail in the joints and segments ensures realistic movement and flexibility, making it an excellent tool for studying prosthetics and robotics. 

\paragraph{{16. 12V Power Supply Cover}}
Designed to serve as a protective cover for a 12-volt power supply. The cover is a rectangular, open-top enclosure that provides protection and organization for the power supply unit. The main body of the cover is a flat, rectangular base with raised sides along the perimeter. The sides are smooth and even, forming a sturdy enclosure that securely holds the power supply in place. The cover’s dimensions are precisely designed to fit the standard size of a 12-volt power supply unit. One of the shorter sides features a rectangular cutout, allowing easy access to the power supply's connectors and controls. This cutout ensures that cables can be connected without obstruction and that the user can easily reach any switches or buttons on the power supply. On one of the longer sides, there is a horizontal slot running almost the entire length of the cover. This slot is designed for ventilation, allowing air to circulate and prevent the power supply from overheating. The slot is wide enough to provide adequate airflow while maintaining the structural integrity of the cover. The other sides of the cover are solid, providing robust protection against dust, debris, and accidental contact. The design ensures that the power supply is securely housed while allowing necessary access and ventilation. 

\paragraph{{17. Low-Poly Cat}}
A cat in a low-poly design, emphasizing geometric shapes and angular facets. The cat is depicted in a relaxed, lying-down position with its body stretched out and its tail curling around its side. The body of the cat is elongated and sleek, composed of various flat surfaces and sharp edges that create a faceted appearance. The legs are tucked underneath the body, adding to the relaxed posture. Each leg is angular, with clear, distinct planes forming the shape of the paws. The head of the cat is proportionately large, with a triangular shape that includes prominent, pointy ears at the top. The ears are sharply angled and rise prominently from the head, adding to the geometric aesthetic. The eyes are represented by indented, flat surfaces, giving the impression of depth and shadow. The nose and mouth are subtle, formed by slight indentations and angular cuts on the face. The tail is long and slender, extending smoothly from the rear of the body and curling around to the side. The tail maintains the low-poly style, with clear, angular bends and sharp edges. 

\paragraph{{18. Seattle Space Needle}}
Depicts the iconic Seattle Space Needle, known for its futuristic design and distinctive silhouette. The structure stands tall with a narrow, tapered shaft that widens near the top to support the observation deck and crown. The base of the model features a broad, circular platform that provides stability and support for the tower. From this base, the tower rises with a slender, tapering shaft composed of four primary support legs. These legs are interconnected with horizontal and diagonal braces, adding structural integrity and a visually intricate design. About two-thirds up the tower, the structure widens into the observation deck area. The observation deck is encircled by a wide, flat ring that extends outward, forming a broad viewing platform. This platform is supported by additional struts and braces that connect it to the main shaft of the tower. The deck itself has a smooth, sleek surface, mimicking the real-life appearance of the Space Needle’s observation area. Above the observation deck, the tower continues to rise into the needle-like spire. This spire is slender and pointed, culminating in a sharp tip that emphasizes the tower’s height and elegance. The spire is an essential part of the tower's iconic profile, adding to the overall futuristic look of the structure. 

\paragraph{{19. Anatomic Human Foot}}
Accurately depicts the anatomical structure of a human foot and the lower extremity. It provides a realistic and detailed view of the bones and joints. The foot is composed of various segments, starting with the toes, which are slender and slightly curved, each consisting of multiple segments. These segments connect to longer bones that extend toward the midfoot, forming the arch. The midfoot is supported by a robust cluster of bones that create the base for the ankle region. These bones are larger and more irregularly shaped than those in the toes and midfoot. Above the midfoot, the ankle region is formed by a combination of bones that connect the foot to the lower leg. This area includes a large, sturdy bone that forms the heel, providing support for weight-bearing. The bones in the lower part of the model extend up to show the connection with the lower leg, illustrating the interaction between the foot and the leg. The model's joints are clearly defined, showing the points of articulation and allowing for an understanding of movement and connectivity within the foot and ankle. The surface of the bones is textured to highlight natural irregularities, enhancing the realistic appearance. Overall, the model is proportionally accurate, ensuring that the structure of the foot and its connection to the lower leg are anatomically correct. Each segment transitions smoothly into the next, providing a comprehensive view of the foot's anatomy and its essential functions. 

\paragraph{{20. Universal Joint}}
Represents a universal joint, a mechanical device that allows two shafts to rotate and transmit torque while allowing for angular misalignment. The universal joint is composed of several interconnected components, each designed to facilitate smooth and flexible movement. The main body of the universal joint consists of two yokes, which are U-shaped components that house the cross-shaped central pivot. Each yoke has a pair of parallel arms extending outward, forming a U-shape. The arms of each yoke are connected to the central pivot through a pin, allowing for rotation. The central pivot, also known as the cross or spider, has four arms arranged in a cross shape. Each arm of the pivot is inserted into the arms of the yokes, creating two perpendicular rotational axes. This configuration allows the universal joint to accommodate angular misalignment between the connected shafts. The model features smooth surfaces and precise dimensions, ensuring that the components fit together seamlessly. The pins connecting the yokes to the pivot are cylindrical and securely fastened, allowing for smooth rotation without any play or looseness. At each end of the yokes, there are cylindrical extensions that represent the shafts connected to the universal joint. These extensions are designed to fit into corresponding openings in the shafts, providing a secure connection and efficient torque transmission. 

\paragraph{21. Bolt}
Depicts a bolt, showcasing a detailed and accurate representation of this essential fastening hardware. The bolt features a cylindrical shaft with helical threads and a hexagonal head for easy tightening and loosening. The shaft of the bolt is smooth and cylindrical, with evenly spaced helical threads running along its entire length. These threads are designed to engage with a corresponding nut or threaded hole, providing a secure and strong fastening. The threads are precise and consistent, ensuring that the bolt can be screwed in smoothly and hold firmly in place. At one end of the shaft is the hexagonal head, which is wider than the shaft and has six flat faces. This hexagonal shape allows for easy gripping with a wrench or socket, making it simple to apply torque to tighten or loosen the bolt. The edges of the hexagon are sharp and well-defined, ensuring a good grip with tools. The end of the shaft opposite the head is flat, providing a smooth surface for the bolt to press against when fully inserted into a threaded hole. This flat end ensures that the bolt can be tightened securely without damaging the surface it is fastened to.

\paragraph{{22. Forbidden Watchtower}}
Depicts a detailed and intricately designed watchtower. The tower stands tall and imposing, with a rugged, medieval appearance that adds to its mystique. The watchtower features a cylindrical structure, tapering slightly as it rises. The base of the tower is surrounded by a rocky terrain, with uneven and jagged stones that provide a natural, rugged foundation. The stones are detailed with textures and variations, giving the base a realistic, weathered look. Ascending from the rocky base, the cylindrical tower is built with stone blocks, each block clearly defined and textured to resemble rough-hewn stone. About halfway up the tower, there is a row of small, narrow windows that encircle the structure, providing a vantage point for lookouts. These windows are evenly spaced and add to the tower's defensive feel. Above the windows, the tower continues to rise, culminating into an empty hole, instead of a roof. Just below this hole, there is a band of decorative stonework, including a circular emblem, which adds an element of detail and intrigue to the tower. Overall, the model is well-proportioned, with the height of the tower and the width of the base balanced to create a stable and visually appealing structure. 

\paragraph{23. Spinal Cord}
Presents a detailed representation of the lumbar region of the human spine, complete with congenital defects. The spine is depicted in a vertical orientation, showing the intricate structure and natural curvature of the lumbar vertebrae. The lumbar spine consists of several large, block-like vertebrae, stacked on top of one another. Each vertebra is characterized by a prominent, rounded body at the front, which serves as the main weight-bearing structure. Extending from the back of each vertebra is a complex arrangement of bony projections, including the spinous processes, which are the most noticeable and point posteriorly, and the transverse processes that extend laterally. Between each vertebra, there are spaces where the intervertebral discs would normally be, providing cushioning and flexibility to the spine. The vertebral bodies and processes are textured with fine details that mimic the natural surface of the bone, including small indentations and protrusions. The model also highlights specific congenital defects, which are anomalies present from birth. These defects are shown as irregular shapes or abnormalities in the vertebrae, such as variations in size, shape, or the presence of additional bony structures. These features may include fused vertebrae, extra or missing processes, or atypical curvatures. At the base of the lumbar spine, the model connects to the sacrum, which is a large, triangular bone that fits between the hip bones. The sacrum is detailed with similar texturing and exhibits a series of fused vertebrae, contributing to the model's overall complexity.

\paragraph{{24. RC Chassis}}
Represents a wheel designed for an RC Rocker-bogie chassis, specifically a mini Martian rover. The wheel is constructed with features that enhance its performance on rough and uneven terrain, typical of Martian landscapes. The wheel is circular and robust, with a solid outer rim that provides durability and stability. The outer surface of the rim features a series of raised, rectangular treads arranged in a staggered pattern. These treads are designed to increase traction and prevent slipping, allowing the rover to navigate challenging surfaces more effectively. Inside the outer rim, the wheel has a series of spokes that extend from the central hub to the rim. The spokes are wide and evenly spaced, providing structural support and distributing the load evenly across the wheel. This design ensures that the wheel can handle the stresses of rough terrain without compromising its integrity. The central hub of the wheel is a solid, cylindrical section that serves as the mounting point for attaching the wheel to the rover's chassis. The hub includes a central hole, which is designed to accommodate an axle or shaft, allowing the wheel to rotate smoothly. The hole is surrounded by a flat surface, ensuring a secure fit with the connecting components. 

\end{document}